\documentclass[12pt,a4paper]{article}

 \usepackage[dvips]{graphics}
 \usepackage{graphicx}
 \usepackage{afterpage}
 \usepackage{enumerate}
\begin{document}

\title{
Spectral effects simulation with two-dimensional \\ magnetohydrodynamic models
of the solar photosphere
  }

 \author{I. N. Atroshchenko and V. A. Sheminova }
 \date{}

 \maketitle
 \thanks{}
\begin{center}
{Main Astronomical Observatory, National Academy of Sciences of Ukraine
\\ Zabolotnoho 27, 03689 Kyiv, Ukraine
}
\end{center}

 \begin{abstract}
To study the structure of spatially unresolved features of the solar photosphere,
we calculated the Stokes profiles of seven photospheric iron lines using
two-dimensional nonstationary MHD models of solar granulation for various amounts
of magnetic flux (0, 10, 20, 30~mT). We investigate variations in the absolute
wavelength shifts and bisectors of the I profiles, as well as variations in the
zero-crossing wavelength shifts, amplitude and area asymmetry of the  V profiles as
functions of magnetic field strength and time. The center-to-limb variations of the
Stokes profiles are analyzed. The iron abundance is found to be 7.57, with the
photosphere inhomogeneities taken into account. Although most of the spectral
effects simulated within the scope of the two-dimensional MHD models are in satisfactory
agreement with observational data, these  models cannot always
give a quantitative agreement. The absolute wavelength shifts of the
Stokes profiles of Fe~II lines calculated with the MHD  models  are substantially smaller than the observed ones.
\end{abstract}

\section{Introduction}

Our knowledge of the structure of small-scale magnetic features on the Sun is based
mainly on spectroscopic observations in polarized light. To obtain the information
from observations, we have to take recourse to laborious simulations of spectral
effects. The simulation is made for such spectral line parameters which depend
primarily on the properties of magnetic elements and can be easily interpreted.
Nowadays we are able to study spatially unresolved magnetic elements with strong
magnetic fields. A technique for the diagnostics of small-scale magnetic features
has been elaborated quite well. Concurrent spectropolarimetric observations of
several hundred lines are successfully carried out. Multidimensional MHD models of
solar granulation are built. All these advances make it possible to calculate the
Stokes profiles of spectral lines, compare them with observations, and gain
information on the internal structure of magnetic flux tubes. The most interesting
object of studies in the interpretation of the Stokes profiles is the V-profile
asymmetry found in 1985 in observations of plages and the quiet Sun network
\cite{21,24}. Up to now, no source has been found for the asymmetry of Stokes
profiles (see overview \cite{19}). While the mechanism of area asymmetry can be
understood in general, the amplitude asymmetry may be caused by various factors
(variations of velocity along and across the line of sight, time variations, etc.),
and so we still do not understand what makes the V-profile amplitudes asymmetric
even at the disk center. That is why any attempt to reproduce the observed
asymmetry using MHD models gives an additional valuable information on the
structure of magnetic elements and their environment. A correct reproduction of the
asymmetry is the most rigorous test for solar granulation models.

In this study we intended to calculate the Stokes profiles of photospheric lines
using 2-D MHD models of the solar photosphere described in our paper \cite{2}, to
study profile variations in time and in space depending on amount of magnetic flux
and position on the solar disk, and thus to find whether such theoretical models
can be used to interpret spectropolarimetric observations.

\section{Model  of the solar photosphere
}

Theoretical 2-D nonstationary MHD models of the solar photosphere \cite{2} are
characterized by a high spatial resolution (the horizontal calculation mesh width
is 15~km), the existence of magnetic flux tubes in the region simulated, as well as
the existence of nonmagnetic regions between the flux tubes -- the granules. These
models allow not only the simulation of spectral effects in the Stokes V profiles,
but the analysis of I profiles as well, and this widens the scope of spectral
analysis. Recall major features of these models.

We commonly assume in 2-D models that magnetic elements are either symmetric from
the viewpoint of parallel transfer or axially symmetric. In the first case they are
magnetic flux slabs, and in the second case they are magnetic flux tubes. Our
models deal with flux slabs , but we call them flux tubes by convention. The
simulated region comprises $256 \times 128$ calculation meshes with a horizontal
and a vertical width of 15~km, it is rectangular in shape, 3840~km wide and 1920~km
high, its upper boundary corresponds to the height $H = 600$~km above the level
$\tau_R = 1$. The region may be considered as 256 vertical columns spaced at
intervals of 15~km parallel to one another and to the line of sight. The location,
thickness, and number of magnetic flux tubes in the simulated region depend on mean
amount of magnetic flux and simulation time. The number of flux tubes and their
width at the level $\tau_5 = 1$ increase with mean magnetic flux. The filling
factor in the region grows with magnetic flux. The temperature in the flux tubes in
the region of formation of spectral lines, i.e., above the level $\tau_5 = 1$,
increases as compared to the surrounding nonmagnetic medium. The intensity of the
continuous radiation emerging from flux tubes also increases. The maximum magnetic
field strength in flux tubes is approximately the same in all models ($B = 200$~mT
at $\tau_5 = 1$). The Wilson depression is 150--200~km.

Comparing the temperature ${<}T{>}$ averaged over space and time for our MHD models and
the temperature $T$ for the quiet photosphere (the empirical homogeneous HOLMU
model), we found that ${<}T{>}$ for a theoretical nonstationary nonmagnetic model was
higher than $T$, the excess being as large as 400 K in the region where weak lines
are formed ($\log\tau_5 = -0.5$) and several tens of degrees in the region where
moderate-strong lines are formed. The space-average temperature for MHD models
slightly diminishes with growing simulation time at a particular mean field
strength. The temperature ${<}T{>}$ diminishes as well with increasing mean field
strength in the simulated region.

We used four model sequences in the Stokes profile calculations, the mean magnetic
field strengths in them were 0, 10, 20, and 30~mT, each sequence had five models
calculated for points of 21, 22, 23, 24, and 25 min in time. In all, we had twenty
2-D models. The model for a specific mean strength (for instance, 10~mT) and a
specific point in time (for instance, 25 min) is designated as MHD-10-25.

\section{Calculating the Stokes profiles
}

The transfer equations for a radiation polarized in a magnetic field were solved
separately for every model column. First the optical depth $\tau_5$ in the
continuum along the line of sight was determined, and then the Stokes profiles were
calculated in the LTE approximation by the procedure described by Beckers \cite{10}
and Landi Degl'Innocenti \cite{16}. The algorithm and program for the Stokes
profile calculations are described in detail by Sheminova \cite{5,6}. The mesh for
solving the equations of transfer along the line of sight was built in the
following manner. The initial mesh point (the point on the upper boundary of
simulated region where the radiation emerges on the solar surface) was found from
the model values of the geometric height $H$, continuous absorption coefficient $\kappa_5$,
and matter density $p$ which correspond to two points in the model beginning from the
upper boundary of the region: $\tau_5 = 1/2(H_2- H_1) [(\kappa_5 \rho)_1 +
(\kappa_5 \rho)_2]$ according to the trapezoidal formula. Then all subsequent
points $\tau_n = \tau_{n-1} + 1/2(H_n- H_{n-1}) [(\kappa_5 \rho)_n + (\kappa_5
\rho)_{n-1}]$. Thus the model values of $H$ were scaled to the $\tau_5$ scale and
then to the $\log\tau_5$ scale. After that, we specified a standard mesh nonuniform
in $\log\tau_5$: $\log\tau_5$ was 0.2, 0.15, 0.1, 0.05, and 0.1 in the intervals
between $\log\tau_5$ values of -6.4, -5.2, -4.0, -3.0, -0.5, and 1.0. All model
parameters were calculated according to the standard mesh, and our experience
suggests that the mesh gives stable results. The number of mesh points did not
exceed 95. The models were calculated for various distances from the disk center
with $\cos\theta$ taken into account. The models were thus made ready for exact
calculations of the transfer equations of Unno-Beckers-Landi Degl'Innocenti.

The Stokes profiles were calculated for each column, then they were averaged over the
simulated region, and we got a profile for a particular magnetic flux and a
particular point in time. As we had five models for points in time 21, 22, 23, 24,
and 25 min, we had five corresponding profiles. We averaged them over time and
found a resulting profile for a specific magnetic flux, and this profile could be
compared with observations which have a low spatial and temporal resolution. To
study the center-to-limb variations in the Stokes profiles, the calculations were
made for five values of $\cos\theta= 0.3$, 0.45, 0.67, 0.83, 1.0. We picked the
optimum horizontal mesh width which would provide the necessary accuracy with
shortest computing time. Calculations of a profile averaged over time and over
columns with mesh widths of 15, 30, 45, 60, and 90~km suggested that the optimum
mesh width was 60~km, and thus we worked with 64 columns rather than with 256.

\subsection{Selected spectral lines }

We selected seven lines which are often used in observations of magnetic features.
Table~1 gives their atomic parameters: the low excitation potential $\chi_e$,
effective Land\'{e} factor $g_{\rm eff}$, oscillator strength $\log gf$ from
\cite{11} (and from \cite{23} for the line  Fe  I $\lambda$~526.06~nm and for Fe~II
lines), effective optical depths of formation of the profile center and the whole
profile -- $\log\tau_d$ and $\log\tau_W$. The damping constant was calculated with
Unsold's formula. The lines were selected so that the photospheric region where
they are formed might be as extended as possible. As is evident from the effective
line formation depths calculated with the depression contribution functions for the
quiet Sun \cite{3}, the lines selected by us refer to that region in the
photosphere which extends from $\log\tau_5=-1$ to $\log\tau_5 = -3$, and therefore
all results of our spectral analyses refer to just this part of the photosphere.
The number of lines was limited mainly by great time expenditure in the Stokes
profile calculations for the MHD models.

%

{\footnotesize
 \begin{table} \centering
 \parbox[b]{15.5cm}{
\caption{Lines used in the analysis
 \label{T:5}}
\vspace{0.3cm}}
 \footnotesize
\begin{tabular}{lccclccc}

 \hline
Ion& $\lambda $,\,nm & $\chi_e$,\,eV & $g_{\rm eff}$ &$\log gf$ &$\log\tau_{d}$&
$\log\tau_W$\\
 \hline
Fe I&  524.70585 &0.09  & 2.00 & -4.946 & -2.65 &  -2.05         \\
Fe I&  525.02171 &0.12  & 3.00 & -4.938 & -2.60 &  -2.02        \\
Fe I&  525.06527 &2.20  & 1.50 & -2.120 & -3.15 &  -2.09        \\
Fe II& 614.92490 &3.89  & 1.33 & -2.850 & -1.05 &  -0.88        \\
Fe I&  615.16220 &2.18  & 1.83 & -3.299 & -1.78 &  -1.50        \\
Fe I&  617.33410 &2.22  & 2.50 & -2.880 & -2.26 &  -1.75        \\
Fe II& 643.26830 &2.89  & 1.83 & -3.760 & -1.22 &  -1.00         \\

    \hline
 \end{tabular}
 \end{table}}
 \noindent

The observational data we used in this study were taken from various investigations
published already. The observed Stokes profiles for three lines obtained with a
high spatial and temporal resolution in plages near the solar disk center were
taken from paper \cite{9}. The entrance slit was 0.5$^{\prime\prime}$ and the
exposure time 0.2--0.5~s in those observations. The center-to-limb variations in
the amplitude, asymmetry, and profile parameters were observed with a
5$^{\prime\prime}$ slit and an exposure time of 43--72~min \cite{22}. These two
observation series differ basically by their spatial and temporal resolution. We
used observations of other authors as well.

\subsection{Calculation scheme }

All parameters in our self-consistent MHD models are interdependent. Chemical
abundance and damping constant are the only free parameters in the profile
calculations. Therefore, the first stage in the calculations was to determine these
parameters for every spectral line by fitting the equivalent widths and/or central
depths of the I profiles calculated with the MHD models with a zero magnetic flux
to those observed for the quiet Sun as given in atlas \cite{13}. The second stage
was to calculate the Stokes profiles for every line with the abundance and damping
constant found for that specific line and with the MHD models with a magnetic
field. Then the profile parameters were calculated -- bisectors, absolute
wavelength shifts, amplitude and area asymmetry, zero-crossing wavelength shifts,
peak separation, amplitude ratios for pairs of lines, etc. The third stage was to
compare the calculation results with observations.

\section{Analysis of the calculated  Stokes I profiles \\ for the quiet Sun
}
\subsection{Iron abundance}

The iron abundance determined by us is given in Table 2, where $d$ and $W$ are the
central depth and the equivalent width of I profiles measured in atlas \cite{13},
$H_d$ and $H_W$ are the effective geometric height of line core formation and the
weighted average effective height over the whole profile calculated in \cite{3} for
the HOLMU model, $A_d$, $A_W$ are the iron abundances obtained by fitting central
depths and equivalent widths; $\Delta A_{W-d}$ is the difference between the  abundances
$A_W$ and $A_d$. The differences $\Delta A$ calculated with the MHD-0 models are in good
agreement with those found in \cite{3} with the HOLMU model, except for strong
lines, for which the difference becomes greater.

%

{\footnotesize
 \begin{table} \centering
 \parbox[b]{15.5cm}{
\caption{Iron abundance
 \label{T:5}}
\vspace{0.3cm}}
 \footnotesize
\begin{tabular}{ccrccccrr}

 \hline
$\lambda $,\,nm & $d$ & $W$,\,pm & $H_W$,\,km &$H_d$,\,km& $A_W$& $A_d$& $\Delta
A_{W-d}$& $\Delta A_{W-d}$ \cite{3}\\
 \hline
 524.70585 &0.716  & 6.036 &328 &415  & 7.48 &  7.65& -0.17&   -0.11        \\
 525.02171 &0.710  & 6.490 &324 &409  & 7.58 &  7.65& -0.07&   -0.03        \\
 525.06527 &0.793  &10.400 &330 &493  & 7.52 &  7.39& -0.13&   -0.08        \\
 614.92490 &0.347  & 4.030 &132 &164  & 7.63 &  7.52&  0.11&    0.10        \\
 615.16220 &0.507  & 4.858 &235 &283  & 7.62 &  7.68& -0.06&   -0.06        \\
 617.33410 &0.622  & 6.930 &279 &361  & 7.55 &  7.58& -0.03&   -0.02        \\
 643.26830 &0.364  & 4.190 &159 &194  & 7.64 &  7.56&  0.08&    0.06        \\
   \hline
 Average   &       &       &    &     & 7.57 &  7.57&      &  \\
     \hline
 \end{tabular}
 \end{table}}
 \noindent

Several factors are responsible for the discrepancy between $A_W$ and $A_d$. The
damping constant may be one of them. To verify this assumption, we introduced a
correction factor $E$ in the calculations ($\gamma = E \gamma_{\rm V dW}$). We
calculated the profiles with various correction values ($E= 1$, 2, 3) and found
that the abundance errors caused by uncertainty in the damping constant were 0.02
dex for our lines, and their effect on $A_W$ and $A_d$ was unimportant. We believe
that temperature is the most probable factor for the differences between $A_W$ and
$A_d$. Gurtovenko and Sheminova \cite{4} demonstrated that these differences could
be successfully used for improving solar model photospheres. Now we try to test the
MHD-0 model using the abundances $A_W$ and $A_d$ obtained. If the model temperature
is lower than the actual temperature in the region of line formation, $W$ and $d$
for moderate Fe~I lines in this model are greater than the observed $W$ and $d$
and, therefore, we get $A_W$ and $A_d$ smaller than the actual iron abundance, and
vice versa, i.e., the following relationships hold true: if $T_{\rm mod} < T$,
$A_d< A$ and $A_W < A$; if $T_{\rm mod} > T$, $A_d > A$ and $A_W > A$. As the
spectral lines have different atomic parameters, they react in different ways on
temperature changes, and so we divided the lines into three groups: moderate-weak
wide lines Fe~II 614.92, 643.26~nm; moderate narrow lines Fe~I 525.02, 524.70,
615.16, 617.33~nm; moderate-strong wide line Fe~I 525.06~nm. According to
quantitative estimates by Sheminova \cite{8}, the same temperature fluctuations
$\Delta T$ produce a weak response in the lines in the first group and a strong
response in the second group, while the lines in the third group demonstrate an
intermediate response. The central depths are more responsive in narrow lines, and
equivalent widths in broad lines. We derived the average iron abundances $A_d =
7.55$, $A_W = 7.64$ from Fe~II lines (the first group), $A_d = 7.64$, $A_W = 7.54$
from Fe~I lines in the second group, and $A_d = 7.39$, $A_W = 7.52$ in the third
group. The results for the first two groups are explicable if we assume that the
average temperature in the simulated region is higher than the actual temperature
and the difference $\Delta T$ need be only 20--50~K in the region of formation of
moderate Fe~I lines and 200--400~K in the region of formation of Fe~II lines. The
strong photospheric line in the third group is formed in the most extended region,
which embraces virtually all photospheric layers. As a result the effective layers
of formation of the line core may lie very close to the upper boundary of MHD
models in those model regions where the absorption increases, and then the
abundance $A_d$ is more distorted than $A_W$ or than $A_d$ from moderate lines. Since
the MHD models are truncated and temperature fluctuations can be quite appreciable
in the upper layers, this seems to be the reason why the abundances $A_d$ were
found smaller than $A_W$.

Exact determination of iron abundance is still an open problem (see review
\cite{15}). Analysis of the $A_W$ and $A_d$ obtained by us indicates that the
average temperature is likely to be slightly overestimated in the MHD models in the
region where moderate Fe I lines are formed and is too high in the region where
Fe~II lines are formed. Calculations and analyses of strong photospheric Fe~I lines
with our MHD models cannot give reliable results because the models are truncated
in the upper photosphere. Nonetheless the average iron abundances $A_W$ and $A_d$
derived from all lines by fitting the calculated central depths and equivalent
widths to observed ones coincide and are 7.57 with rms errors of $\pm0.06$ for
$A_W$ and ±0.12 for $A_d$. This iron abundance is close to that in meteorites
(7.55).

\subsection{Absolute wavelength shifts of I profiles }

Absolute Doppler shifts $\Delta\lambda = c(\lambda_c - \lambda)/\lambda$ calculated
for I profiles are given in Table 3. We found the central wavelength $\lambda_c$
for a displaced profile as the center of gravity of the area of I-profile core, for
the profile section cut off by a line at half-maximum intensity. The wavelength
$\lambda_c$ was found to be very sensitive to the method of its determination and
to the wavelength step $\Delta\lambda$ within the profile. Table 3 gives the shifts
$\Delta\lambda$ calculated from profiles with the steps $\Delta\lambda = 0.4$ pm
and $\Delta\lambda = 0.1$ pm. We found that the Doppler shift could not be
determined more accurate than to  0.1 pm, and this results in an error of 50~m/s.
The observed absolute shifts $\Delta\lambda$ are taken from study \cite{14}. It is
well known that line shifts are caused by convective motions of matter. Negative
shifts point to upward motions toward the observer (blue shifts), positive shifts
suggest that the matter moves downward from the observer (red shifts). The absolute
line shifts calculated with our MHD models are too small as compared to
observations and are almost the same.

\subsection{Bisectors of I profiles }

Bisectors are usually used to indicate the asymmetry of I profiles. The absolute
shift of bisectors characterizes the velocity amplitude, and their bend
characterizes the velocity gradient. We have to keep in mind, however, that the
bisector shape is affected by temperature structure changes to a greater extent
than by velocity (see review \cite{1}). The results obtained in our study are
illustrated by bisectors for three lines in Fig.~1. All the bisectors are C-shaped.
The greatest blue bend in the bisectors of stronger lines is smaller than the
observed bend. The calculated bisectors are shifted only slightly with respect to
the observed ones.

%

{\footnotesize
 \begin{table} \centering
 \parbox[b]{15.5cm}{
\caption {Doppler shift (in units of~m/s)
 \label{T:3}}
\vspace{0.3cm}
}

 \footnotesize

\begin{tabular}{ccrr|rcccccc}
 \hline
 \multicolumn{4}{c}   { Quiet Sun} &        \multicolumn{7}{|c} { Magnetic regions}  \\
 \hline
  \multicolumn{4}{c|} { $\Delta\lambda_I$}
& \multicolumn{3}{c|}  { $\Delta\lambda_I$ } &
  \multicolumn{4}{c}  {$\Delta\lambda_V$} \\
 \hline
 $\lambda $,\,nm & 0.4,\,pm &  0.1,\,pm  & \cite{14}&10& 20&30,\,mT& 10& 20&30,\,mT& \cite{9}\\
 \hline
524.70  &   -143  &   57&  -81   &   76 &    145 &    180&     560 &    280  &   116  &   --       \\
525.02  &   -137  &   63&  -133  &   59 &    131 &    171&     548 &    319  &   162  &   --      \\
525.06  &   -120  &   91&  -179  &  131 &    221 &    245&     471 &    365  &   211  &   --     \\
614.92  &   -195  &   5 &  -1108 &  -88 &    84  &    134&    1400 &    496  &   302  &   495    \\
615.16  &   -146  &   19&  -387  &   15 &    95  &    138&     860 &    371  &   171  &   530    \\
617.33  &   -146  &   -9&  -203  &   38 &    129 &    163&     703 &    414  &   231  &   420    \\
643.26  &   -140  &   19&  -491  &  -57 &    103 &    143&    1225 &    575  &   282  &   --      \\
 \hline
Average &   -146  &   35&  --     &   25 &    129 &    147&     824 &    482  &   218  &   --      \\
     \hline
 \end{tabular}
 \end{table}}
 \noindent

\begin{figure}[t]
   \centering
   \includegraphics[width=10.cm]{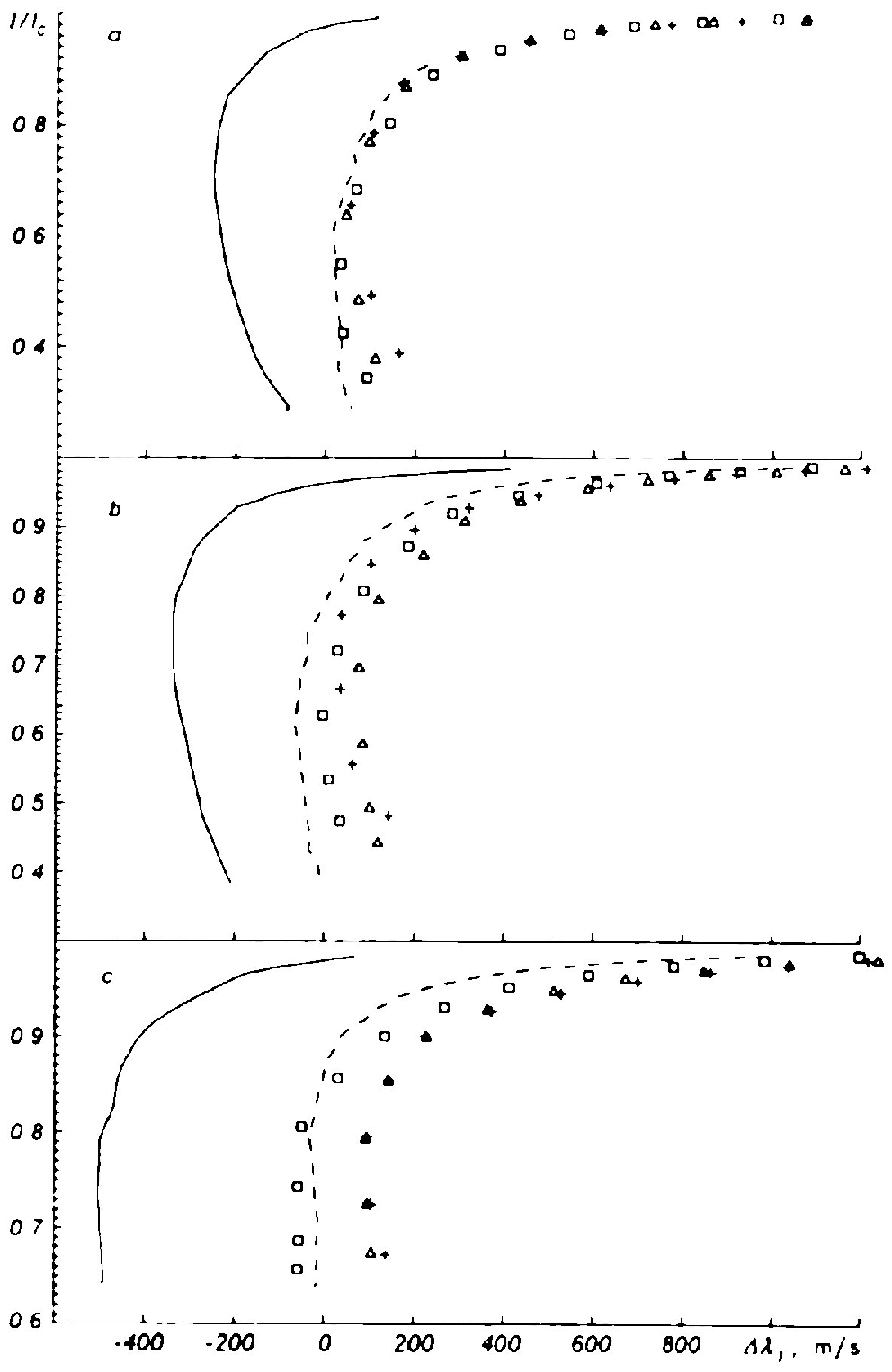}
 \hfill
\parbox[b]{5.3cm}{ \vspace{0.0cm}
{Fig 1. I-profile bisectors on the absolute-shift scale; a) Fe I $\lambda $~524.7~nm, b)
Fe I $\lambda$~617.3~nm, c) Fe II $\lambda$~643.2~nm. Solid line) observations, dashed line)
MHD-0, squares) MHD-10, triangles) MHD-20, plusses) MHD-30.
 } \label{Fig:Fig1}
   }
\end{figure}
\begin{figure}[t]
   \centering
   \includegraphics[width=6.5cm]{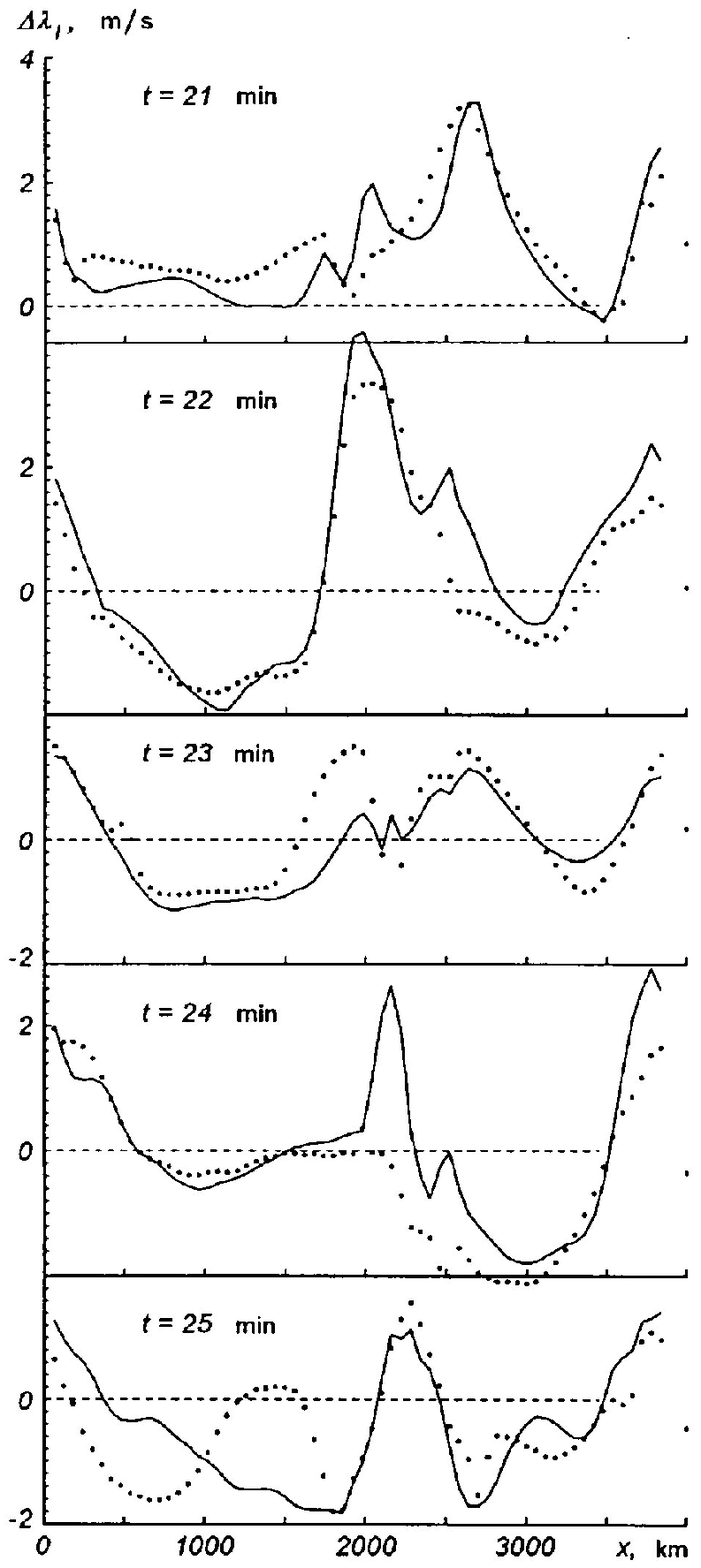}
 \hfill
\parbox[b]{8 cm}{ \vspace{0.0cm}
{Fig.2. Variations in the absolute shifts along the length of the region simulated
with a zero field strength (model MHD-0). Solid line) $\lambda$~614.9~nm, dots)
$\lambda$~525.02 nm.
 } \label{Fig:Fig2}
   }
\end{figure}

So, the calculations of I profiles and their  parameters reveal that MHD models for
the quiet Sun describe adequately the observed spectral effects and reproduce the
well-known relationships, but a quantitative checking suggests that the model is
slightly "overheated", and this is likely to be the cause of small absolute shifts
derived by us. To test this assumption, we studied the variations of the calculated
central depths $d_c = 1 - I/I_c$, equivalent widths $W$, and absolute shifts
$\Delta\lambda_I $ along the region simulated with a zero field strength for points
in time 21--25 min. We selected two lines, $\lambda$~614.9~nm ($H_d \approx
150$~km) and  $\lambda$~525.02~nm ($H_d \approx 400$~km), in order to follow the
parameter variations at different photospheric levels. The line $\lambda$~525.02~nm
being highly sensitive to temperature, the dependence of its central depth on
distance $x$ along the region has sharp dips at the boundaries of granules and
intergranular lanes, where the temperature rises. The equivalent widths follow in
general the run of $d_c(x)$, but there are some peculiarities. The relations
$d_c(x)$ and $W(x)$ account for temperature variations in the region, and
$\Delta\lambda_1(x)$ accounts in full measure for variations in vertical velocity.
Figure 2 displays the absolute shifts along the length of the region simulated by
MHD-0. The shift $\Delta\lambda_1$ is predominantly positive in intergranular lanes
and negative in granules, thus allowing a clear delineation between granules and
integranular lanes. We observed no significant difference in the changes of
$\Delta\lambda_1(x)$ for both lines at different photospheric levels. Oscillations
of vertical velocities in time can be seen in granules, but we could not detect
5-min oscillations, probably because of a long temporal step (1 min). The absolute
shifts calculated for the profiles averaged over the region at the moments 21, 22,
23, 24, 25 min are 517, 5, -51, -127, -316~m/s ($\lambda$~614.9~nm) and 794, -167,
55, -161, -347~m/s ($\lambda$~525.02~nm), and for the profiles averaged also in
time they were 4~m/s ($\lambda$~614.9~nm) and 60~m/s ($\lambda$~525.02~nm). These
shifts are substantially smaller than the observed shifts, especially for the line
Fe~II $\lambda$~614.9~nm. A shift of -133~m/s observed in $\lambda$~525.02~nm is
attained in the calculations at some moments, but a shift of -1108~m/s observed in
the $\lambda$~614.9~nm line has not been reached at all. We may infer that the time
averaging is a very delicate element in simulating absolute shifts. It has to be
done either over a long time interval or with a small time step. There may exist
another cause of small shifts in the region of Fe~II line formation. Observations
indicate that the pattern of convective motions changes in the middle photosphere,
the correlation between brightness and velocity is violated. Blue absolute shifts
become very small at a height of about 300~km and become even red above this level.
Below, at a height of 100~km, blue shifts increase up to 1.5--2~km/s. It is likely
that our models simulate inadequately the violation of the brightness--velocity
relationship, and so the gradient of vertical velocities in the models is too low.
The cause appears to be a too large line opacity coefficient, which results in a
slower cooling of the matter than in the actual solar photosphere.

\section{Analysis of the Stokes profiles calculated \\ for active features
}

Profiles of the Stokes parameters I, Q, U, and V offer considerable scope for
studying physical conditions in small-scale magnetic features. The diagnostics of
magnetic fields is made primarily with the Q, U, and V profiles of some infrared
lines \cite{17} and V profiles of lines in the visible region, and with the use of
the method of V-profile amplitude ratios for pairs of lines \cite{20} or
center-to-limb variations in the V-profile parameters \cite{22}. The diagnostics of
velocities in magnetic elements is made by measuring zero-crossing shifts,
broadening and asymmetry of V profiles. The zero-crossing shift  $\Delta\lambda_V$
is measured with respect to the wavelength of the core center of I profile
($\lambda_c$), i.e., $\Delta\lambda_V =  \lambda_V - \lambda_c$, $\lambda_V$ being
the wavelength of the V-profile center. The V-profile center is a point in the
profile where it crosses the $\lambda$-axis, passing from its blue wing to the red
wing. The wavelength of the I-profile center $\lambda_c$ is  defined as a point
where $d(I/I_c)/d\lambda = 0$ rather than the center of gravity of the profile
core. The V-profile broadening is defined by the full width at half-maximum (FWHM)
of the blue and red wings and by the difference $\Delta\lambda_a$ between the blue
and red wing maxima. The V-profile asymmetry is measured by the parameters $\delta
a_V = (a_b - a_r) / (a_b + a_r)$ and $\delta A_V = (A_b - A_r) / (A_b + A_r)$, $a$
and $A$ being the amplitude and area of the blue ($b$) and red ($r$) wings (they
are usually taken without sign). The temperature diagnostics in spatially
unresolved magnetic elements is done from the Q, U, and V profiles of strong lines
and also with the use of the V-profile amplitude ratio between two lines with
different temperature response.

\subsection{The Stokes I and V profiles }

Observed I and V profiles (with large $a_b,~a_r$) can be found in paper \cite{9},
they were obtained for three lines (Fe~I 615.1~nm, 617.3~nm, Fe~II 614.9~nm) in the
plage regions 0.63$^{\prime\prime}$ in size, the observation time was 0.5 s. The
lines Fe~II 614.9~nm and Fe~I 615.1~nm were observed simultaneously in an active
plage near a spot at $\cos\theta = 0.953$, and the line Fe~I 617.3~nm was observed
in another intense plage near a spot at $\cos\theta = 0.866$. We compared these
profiles with calculated profiles averaged over the entire simulated region (about
7$^{\prime\prime}$) and over time (5 min).  The mean field strength in this region
was 30~mT.
\begin{figure}[t]
   \centering
   \includegraphics[width=6. cm]{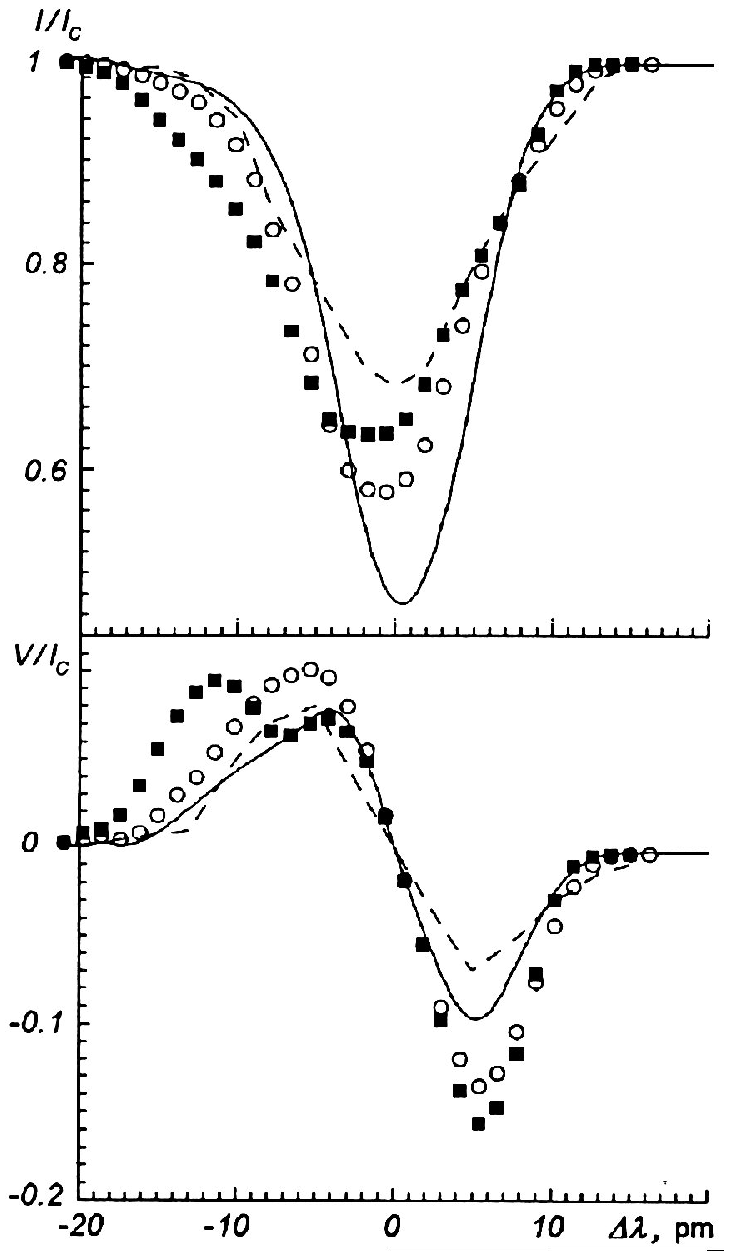}
 \hfill
\parbox[b]{8 cm}{ \vspace{0.0cm}
{Fig.~3. The Stokes I  and V profiles of the line Fe~I $\lambda$~617.33~nm
calculated for the MHD-30 model. Solid line) entire region simulated, circles)
magnetic flux tube region 1800--2800~km, dark squares) magnetic flux tube region
2200--2600~km, dashed line) observations with high spatial resolution \cite{9}.
 } \label{Fig:Fig3}
   }
\end{figure}
No Stokes profiles calculated with our models could fit the observed
profiles. Then we calculated anew the profiles for the Fe~I $\lambda$~617.3~nm line,
having changed the spatial averaging. We selected two areas 1.5$^{\prime\prime}$ and
0.6$^{\prime\prime}$ in size from the simulated region with a stronger magnetic flux
tube located at the center of these areas (region 2 in Fig.~3 \cite{2}). Thus we
increased substantially the filling factor. Figure 3 displays the I and V profiles
averaged over time for the entire region (0--3840~km), for the entire region 2
(1800--2800~km), and for the central part of region 2 (2200--2600~km). The results
of our calculations confirmed the assumption of Amer and Kneer \cite{9} that the
filling factor is of great importance for the cases considered. Time variations in
the profile do not reproduce the observed weakening of I profiles, while spatial
changes of the region seem to account for the observed effect. Notice that the
$\Delta\lambda$ zero-point in Fig.~3 coincides with the zero-crossing wavelength
$\lambda_V$ for the V profiles, just as this was done in \cite{9}.

\begin{figure}[t]
   \centering
   \includegraphics[width=8.5cm]{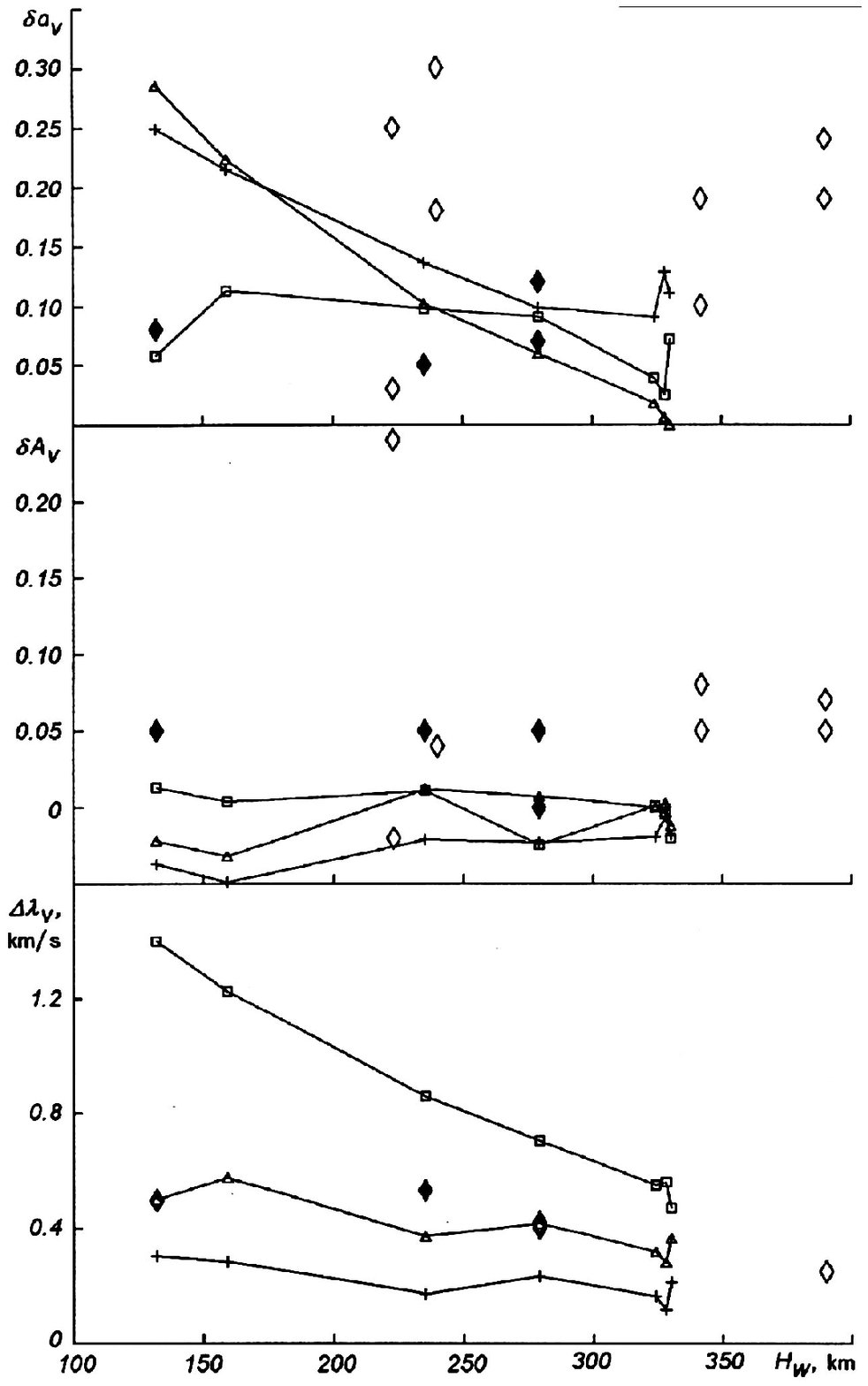}
 \hfill
\parbox[b]{6 cm}{ \vspace{0.0cm}
{Fig.~4. Amplitude asymmetry $\delta a_V$ and area asymmetry $\delta A_V$, and
zero-crossing wavelength shift $\Delta\lambda_V$ as functions of  formation height
for seven iron lines  with the models MHD-10 (plusses), MHD-20 (triangles), and
MHD-30 (squares). Dark diamonds) high-resolution observations \cite{9}, light
diamonds) low-resolution observations \cite{18}.
 } \label{Fig:Fig4}
   }
\end{figure}
\begin{figure}[t]
   \includegraphics[width=8.5cm]{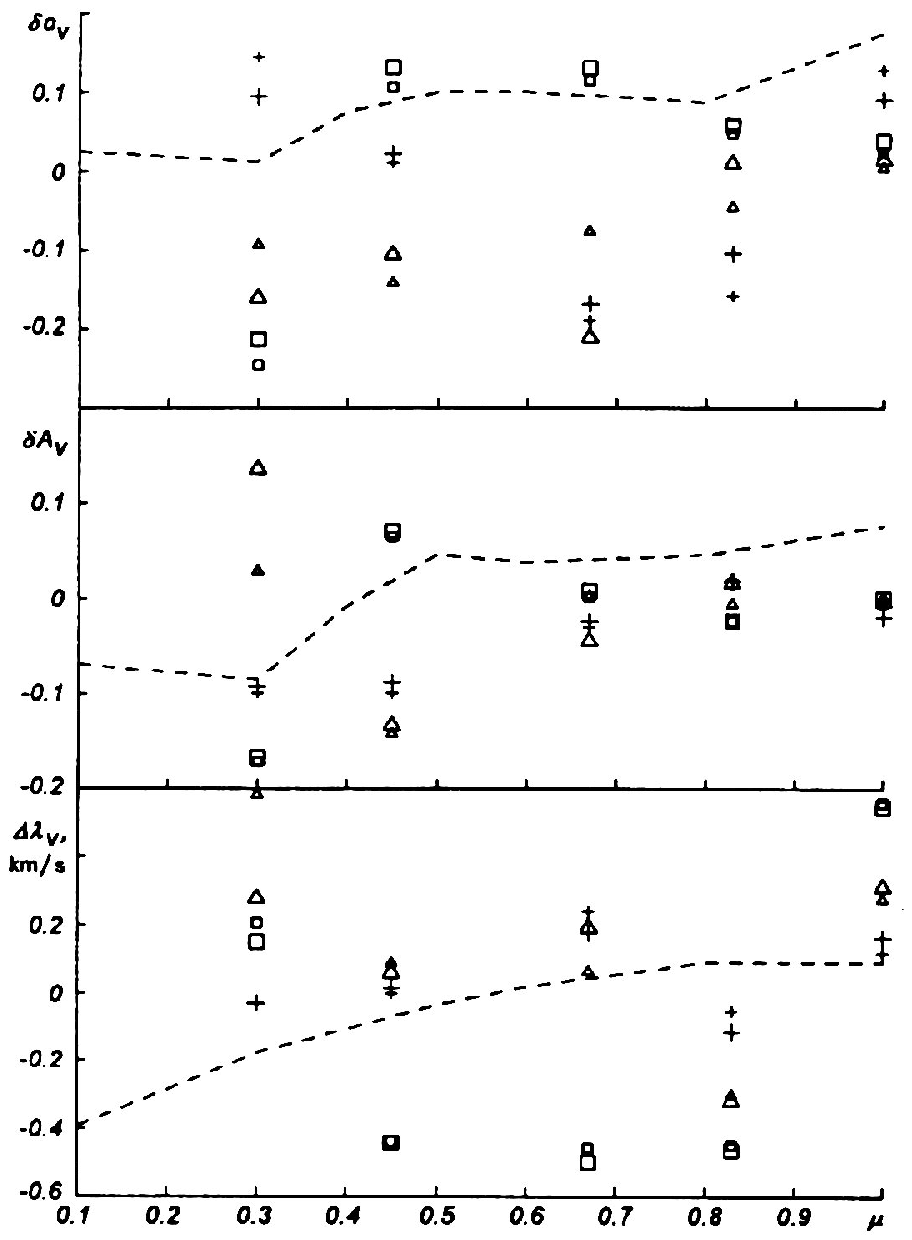}
  \hfill
\parbox[b]{6. cm}{ \vspace{0.cm}
{ Fig.~5. Center-to-limb variations in the amplitude asymmetry $\delta a_V$, area
asymmetry $\delta A_V$, and zero-crossing shift $\Delta\lambda_V$ for MHD-10
(squares), MHD-20 (triangles), MHD-30 (plusses). Larger marks) line
$\lambda$~525.02~nm, smaller marks) $\lambda$~524.7~nm.
 } \label{Fig:Fig5}
   }
\end{figure}

\subsection{ Amplitude  and area asymmetry  and  zero-crossing \\
wavelength shifts of the V profiles}

Figure 4 shows  the asymmetry of amplitudes ($\delta a_V$) and areas ($\delta A_V$)
as well as zero-crossing wavelength shifts  ($\Delta\lambda_V$) derived for  V
profile calculated with MHD models and observed  taken from the literature. The
calculated profiles were obtained for three regions at the disk center with mean
field strengths of 10, 20, and 30~mT, the time interval was 21--25~min.  The
effective heights $H_W$ of formation of each V profile  were taken from Table~2.
Our calculations \cite{7} indicate that the heights $H_W$ calculated with a
magnetic field and without it are practically the same. Besides, the effective
heights for I and V profiles differ insignificantly. As can seen from Fig.~4, the
calculated amplitude asymmetry fits quite well the measurements  in plages with
high spatial resolution \cite{9}, while the calculated area asymmetry is
systematically smaller than the observed one. The area asymmetry varies from 0 to
0.08 as measured by various observers and from -0.05 to 0.02 in our calculations,
i.e., our models do not reproduce  the observed area asymmetry $\delta A_V$. The
zero-crossing shifts $\Delta\lambda_V$ derived for simulated regions with a mean
field strength of 20~mT are in accord with observations made with a high spatial
resolution \cite{9}. As regards the absolute shifts of I profiles (see Table 2),
our calculations do not reveal any blue shift as we have already noted for the
region with a zero magnetic flux. The absolute shifts $\Delta\lambda_I$, averaged
over all lines were found to be 35, 25, 129, 147~m/s for mean magnetic fluxes of 0,
10, 20, 30~mT, respectively. One can see that the effect of shift "reddening" with
growing filling factor does really exist. Brandt and Solanki \cite{12} measured
$\Delta\lambda_I =$~-223, -125, -104, -164~m/s for different filling factors --
$\alpha < 1$\%, 1\%~$\leq \alpha <$~4\%, 4\%~$ \leq\alpha <$~8\%, or $\alpha
\geq$~8\% respectively. Thus, we may infer that the simulated asymmetry and shift
parameters follow the general trends observed in the behavior of these parameters.
The calculated V-profile amplitude asymmetry and zero-crossing  shift fit
observations. The fact that we derived underestimated parameters for the V-profile
area asymmetry and absolute shifts of I profiles suggests that the velocity field
structure in the MHD models  \cite{2} is inadequate.

Figure 5 illustrates the behavior of the calculated amplitude asymmetry $\delta
a_V$, area asymmetry $\delta A_V$, and zero-crossing shift $\Delta\lambda_V$ as
functions of $\mu$, the position of the simulated region on the disk.
Low-resolution observations of V profiles of the lines Fe~I~524.7, 525.02~nm
\cite{22} are shown by dashed lines, the data were averaged over many plages which
differed by their filling factors. Our calculations were made for three regions
with 10, 20, and 30~mT in the time interval 21--25 min, for different values of
$\mu$. The calculated dependence of $\delta a_V$, $\delta A_V$, $\Delta\lambda_V$,
agrees satisfactorily with the observed one. For $\mu=1$  the asymmetry parameters
of both lines  Fe~I~524.7, 525.02~nm are smaller than the observed asymmetry
parameters.

\begin{figure}[t]
   \includegraphics[width=8.5cm]{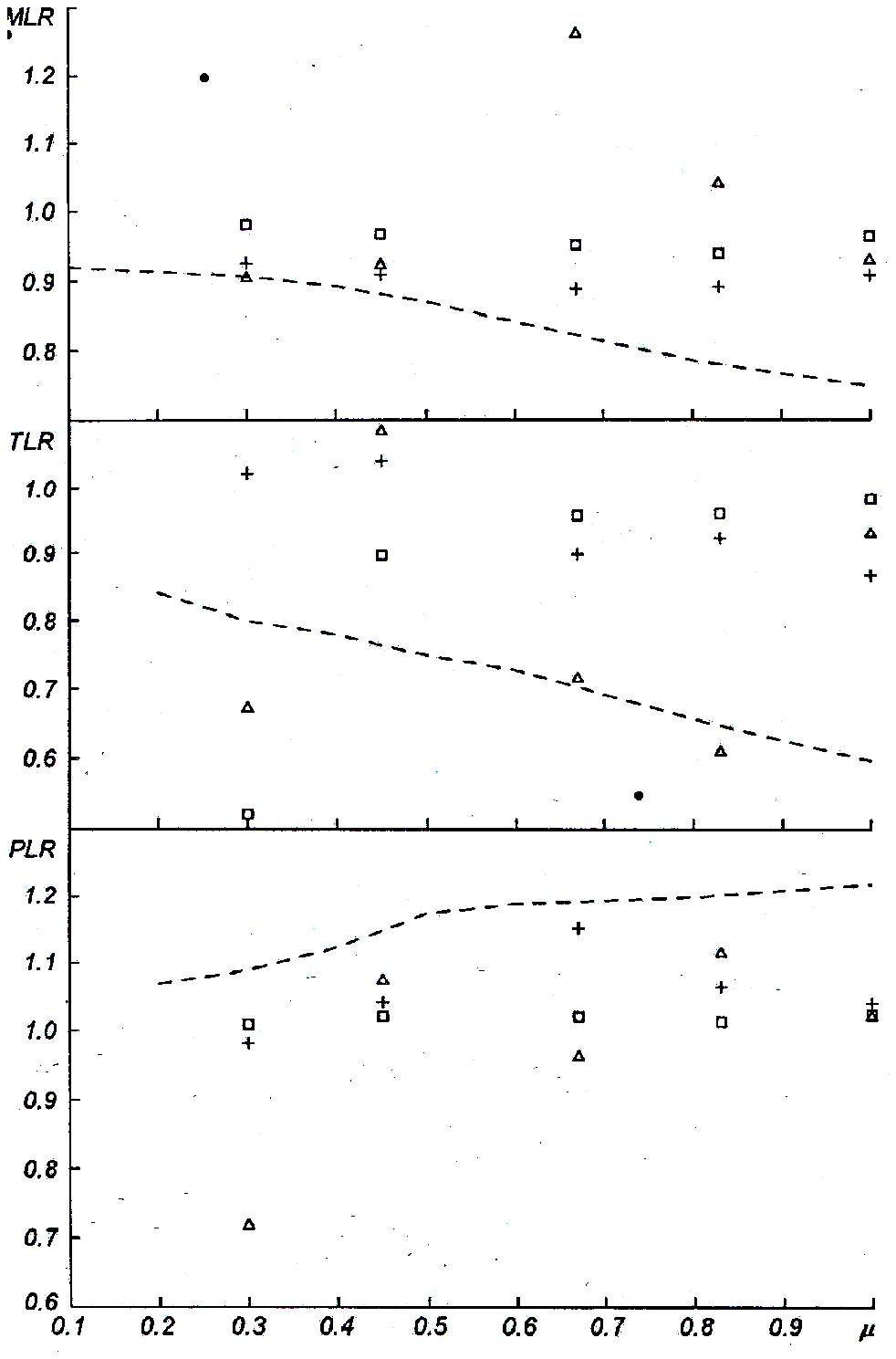}
 \hfill
\parbox[b]{6.5 cm}{ \vspace{0.0cm}
{Fig.~6. The  Stokes  V-amplitude magnetic (MLR),  temperature (TLR) line ratios,
and the Stokes V peak separation line ratio (PLR) as functions of position on the
solar disk ($\mu$) for MHD-10 (squares), MHD-20 (triangles), MHD-30 (plusses).
Dashed line represents observations \cite{22}.

 } \label{Fig:Fig6}
   }
\end{figure}

\subsection{Line ratios }

Figure 6 shows the magnetic line ratios MLR =~$a_{V,\,525.02}/1.5a_{V,\,524.70}$,
the Stokes V peak separation line ratios PLR =~$\Delta\lambda_{a,\,525.02}/
\Delta\lambda_{a,\,524.72}$, and the temperature line ratios TLR =~
$a_{V,\,524.70}/0.75a_{V,\,525.06}$ calculated by us for different values of $\mu$,
the observed ratios \cite{22} are also given. The relations MLR$(\mu)$ and
PLR$(\mu)$, which provide information on the magnetic field gradient, indicate that
the calculated ratios MLR are higher than the observed ones, especially at the
center of the disk. The ratios must grow toward the limb if $B$ decreases with
height, but our data suggest that this effect is weak. It is likely that the field
gradient in the flux tubes simulated is lower than in actual flux tubes in the
region of formation of these lines. The calculated PLR values are smaller than the
observed ones: the effect is the same as for the V-profile amplitude ratios
discussed above.

To study the temperature stratification in small-scale magnetic elements, we
selected the lines Fe~I~525.06, 524.7~nm. The first line being stronger than the
second one, $a_{V\,525.06}$ for it is smaller by 13\% as compared to
$a_{V,\,524.70}$, and therefore ${\rm TLR} \approx 1.15$ at the center of the disk
and ${\rm TLR} \approx 1$ at the limb. Deviations from these values specify the
temperature variation in flux tubes. Observations \cite{22} give the value ${\rm
TLR} < 1.15$. This means that the line becomes weaker in a flux tube owing to a
temperature increase with respect to a nonmagnetic environment. The deviation from
unity diminishes toward the limb -- the temperature difference $\Delta T$ between
the flux tube and the medium decreases. As can be seen in Fig.~6, the calculated
values of TLR are higher than the observed ones. This suggests that $\Delta T$ in
the simulated areas are smaller as compared to actual temperature conditions
observed in plages.

The results obtained with the line-ratio method indicate that the magnetic field
strength gradients and the temperature distribution in flux tubes are not
reproduced adequately by our MHD models. Observations require larger gradients and
higher temperatures.

\section{Conclusion
}

Simulation of spectral effects in the Stokes profiles with the use of
multidimensional MHD models demonstrates once again that construction of such
models is of prime importance for the interpretation of spectra with the aim to
study the structure of the atmospheres of the Sun and stars. The simulation of
spectral effects in V profiles carried out earlier with 1-D models of magnetic flux
tubes proved to be incapable of adequate representation of observations. Asymmetry
is not reproduced in the one-dimensional approximation (e.g., see \cite{18}). The
two-dimensional simulation employed in this study improved considerably the
qualitative results, though it is still far from the desired performance. The
principal results of this study are as follows.

\begin{enumerate}

 \item The iron abundance derived from the equivalent widths and central depths of
selected spectral lines is $7.57\pm0.10$.

 \item The bisectors of I profiles are C-shaped. The blue bend of bisectors in
moderate-strong lines is smaller than the observed one.

 \item We obtained smaller absolute wavelength shifts, especially for Fe~II lines. For
instance, the observed shifts $\Delta\lambda_I$ for the lines Fe~I
$\lambda$~525.02~nm, Fe~II $\lambda$~614.9~nm are -133 and -1108~m/s, while the
calculated shifts are 63 and 5~m/s. The calculated absolute shifts ``redden'' with
growing mean field strength in the region simulated, just as the observed shifts
do.

 \item The amplitude asymmetries calculated for the V profiles of the lines
studied, $\delta a_V$, fit satisfactorily the observations made with high temporal
and spatial resolution, they range from 0 to 0.3. The deeper the line is formed,
the greater is the asymmetry $\delta a_V$, while no direct dependence on magnetic
field strength is found.

 \item The  asymmetry of V-profile areas was found to be smaller than the observed
one -- our calculations give $\delta A_V$ from -0.05 to 0.02, while the observed
values range from 0 to 0.08. The asymmetry $\delta A_V$ grows with field strength
and is virtually independent of the line formation height.

\item The zero-crossing shifts in V profiles  relative to I profile are larger
for the lines which form deeper in the photosphere, they also increase with
magnetic flux. For regions simulated with mean field strengths of 10, 20, 30~mT we
derived shifts (averaged over all lines) of 824, 482, 218~m/s, respectively, while
low-resolution observations in the plages usually produce shifts of about 200~m/s
and high-resolution observations in the plages give about 400~m/s. The agreement
may be considered as quite satisfactory.

\item The Stokes profiles as well as their temperature and magnetic ratios calculated as
function of height in the photosphere and position on the solar disk reproduce
qualitatively the tendencies observed in the behavior of these parameters, but the
quantitative estimates are not consistently in agreement.
\end{enumerate}

Our final inference is that our models are ``overheated'' in the upper photosphere
and as a consequence the gradients of velocities and magnetic field are
underestimated. We believe that the overheating is caused by a too high opacity
coefficient in lines, and in addition, the simulation results for the upper layers
are strongly affected by boundary conditions. Thus, the synthesis of spectral
effects is a rather rigorous test for MHD models. On the other hand, further
improvement of numerical simulation methods advances our understanding of the
physical processes responsible for the formation of small-scale magnetic elements
and the observational spectral effects related to them.

{\bf Acknowledgements.} We thank A. S. Gadun for useful comments and advice. The
study was partially financed by the Joint Foundation of the Government of Ukraine
and the International Science Foundation (Grant No. K111100) and by the European
Southern Observatory (Grant No. A-01-009).



\end{document}